# Tools for Terminology Processing

C. Enguehard*, B. Daille*, E. Morin*

\* IRIN, Université de Nantes,
2, chemin de la Houssinière B.P. 92208, 44 322 Nantes cedex 03, France
{Enguehard, Daille, Morin}@irin.univ-nantes.fr

phone : 33 (0)2 51 12 58 55
fax : 33 (0)2 51 12 58 12

## Abstract

Automatic terminology processing appeared 10 years ago when electronic corpora became widely available. Such processing may be statistically or linguistically based and produces terminology resources that can be used in a number of applications : indexing, information retrieval, technology watch, etc.

We present the tools that have been developed in the IRIN Institute. They all take as input texts (or collection of texts) and reflect different states of terminology processing: term acquisition, term recognition and term structuring.

Key-words : Terminology processing, term acquisition, term recognition, term structuring

# INTRODUCTION

The terminology of a domain is an important resource for researchers from different areas: lexicographers, translators, and also domain specialists (engineers, technicians, redactors of technical documentation, etc.). In addition, specialists in Natural Language Processing have realized that this knowledge could be useful for a numbes of other applications, such as indexing, automatic or machine-aided translation, text generation, semi-automatic abstraction, edition of hypertext systems, ontology building, etc.

However, collections of terms are rare because they are difficult to constitute: terminologists have to study large amounts of documentation about the targeted domain to extract the terms, give their definition, and structure them following a hierarchical classification.

In the 90's, the compilation of large electronic corpora made it possible to build systems to carry out different tasks in terminology processing automatically, especially terminology acquisition, terminology recognition, and terminology structuring.

We present three tools which can assist in these different terminology processing tasks, and we observe their capability of multilingual terminology processing.

# 1. TERM ACQUISITION

Many different professional groups are interested in the automatic acquisition of terminology through corpus exploitation to achieve a number of goals: building of glossaries, vocabularies and terminological dictionaries, text indexing, automatic translation, building of knowledge databases, edition of hypertext systems, etc. The appearance of large electronic corpora in the 90s rendered the development of terminology extractors which can assist terminological work.

Different approaches arose: linguistic (CLARIT (ref. 1), LEXTER (ref. 2), TERMS (ref. 3)), heuristic or statistical (ANA (ref. 4)), or hybrid linguistic-statistical (ACABIT (ref. 5), NEURAL (ref. 6).

The systems we present analyse electronic texts issued from a specialised domain and produce a list of candidate-terms. These candidate-terms have to be validated by a terminologist or a specialist of the domain).

These systems are especially designed for two tasks:

*Terminology mining*

Terminology mining refers to several functionalities:
- To propose a list of candidate-terms ranked from the most representative of the corpus to the least. This list could be used to build specialised dictionaries relative to a new domain;
- To propose a list of candidate-terms for which a morphological variation has been encountered and thus reflect a more advance lexicalisation. This list could be added to update an existing reference list.

*Automatic Indexing*

Automatic Indexing is a subtask of information retrieval which assigns an indexing language to texts (ref. 7). This language consists of a set of descriptors which must reflect the textual document in a discriminatory and a non-ambiguous way. There are two modes of indexing:
- Controlled indexing where indexes are occurrences of terms from a controlled vocabulary;
- Free Indexing where indexes are chosen without consideration for an authority list.

With free indexing, the descriptors, which are simple words, are problematic because of their ambiguity (ref. 8), contrary, key-phrase indexes are not ambiguous, but pose problems because of their variations (ref. 9).

## 1.1. ACABIT

ACABIT is a term extractor which takes as input a tagged corpus with part of speech and lemma which has been structured with XML tags to identify the main parts of the text: title, abstract and sentences. ACABIT proposes as output a rank list of multi-word terms for each text.

Candidate-terms which are extracted from the corpus belong to a special type of co-occurrence:
▪ the co-occurrence is oriented and follows the linear order of the text;
▪ it is composed of two lexical units which do not belong to the class of functional words such as prepositions, articles, etc.;
▪ it matches one of the morphosyntactic patterns of what we will call "base terms", or one of their possible variations.

The patterns for base terms are:
▪ [Noun1 Adj]
    *emballage biodégradable*          *(biodegradable package)*
▪ [Noun1 Noun2]
    *ions calcium*
▪ [Noun1 (Prep (Det)) Noun2]
    *protéine de poissons*           *(fish protein)*
    *chimioprophylaxie au rifampine*    *(rifampicin chemoprophylaxis)*
▪ [Noun1 à Vinf]
    *viandes à griller*           *(grill meat)*

These base structures are not frozen structures and accept several variations. Those which are taken into account are:
▪ Inflexional and Internal morphosyntactic variants:
  — graphic and orthographic variants which gather together predictable inflexional variants:
    *conservation de produit*         *(product preservation)*
    *conservations de produit*       *(product preservation)*
  or not:
    *conservation de produits (products preservation)*
  and case differences.
  — variations of the preposition:
    *chromatographie en colonne*      *(column chromatography)*
    *chromatographie sur colonne*     *(chromatography on column)*
  — optional character of the preposition and of the article:
    *fixation azote*           *(nitrogen fixation)*
    *fixation d'azote*          *(fixation of nitrogen)*
    *fixation de l'azote*        *(fixation of the nitrogen)*

▪ Internal modification variants: insertion inside the base-term structure of a modifier such as the adjective inside the Noun1 (Prep (Det)) Noun2 structure:
    *lait de brebis*           *(goat's milk)*
    *lait cru de brebis*        *(milk straight from the goat)*
▪ Coordinational variants: coordination of base term structures:
    *alimentation humaine*       *(human diet)*
    *alimentation animale et humaine*   *(human and animal diet)*
▪ Predicative variants: the predicative role of the adjective:
    *pectine méthylée*         *(methylate pectin)*
    *ces pectines sont méthylées*    *(these pectins are methylated)*
▪ Morphosyntactic variants: derivational variation that keeps the synonymy of the base term:
    *acidité du sang*          *(acidity of the blood)}*
    *acidité sanguine*         *(blood acidity)*

The program scans the corpus, counts and extracts collocations whose syntax characterizes base-terms or one of their variants. This is done with shallow parsing using local grammars based on regular expressions (ref. 10). These grammars use the morphosyntactic information associated with the words of the corpus after tagging. The different occurrences are grouped as pairs formed by lemmas or morphologically derived lemmas obtained though a morphological analysis of the candidate-term.

These pairs are sorted following an association measure which takes into account the frequence of the co-occurrences.

In addition, ACABIT proposes semantic classes of candidate-terms based on the study of relational adjectives.

## 1.2. ANA

ANA is a term extractor which takes as input a raw data corpus and proposes as output a rank list of terms for each text or for the entire corpus.

This system has been specially design to treat bad quality corpora (for instance written in telegraphic style) which can not be tagged. It does not make any hypothesis on the form of terms.

This incremental system uses a bootstrap (a few frequent terms), a list of functional words, and a list of lexical scheme words.

The system
1. recognises the terms in the texts (and the candidate-terms) included in the bootstrap,
2. collects the context of these recognized terms (through a window of n words),
3. studies these contexts to infer some candidate-terms which will be included in the bootstrap,
4. returns to step 1 unless a stop condition is filled (for instance, there is no more discovery of candidate-terms).

The inference of new candidate-terms follows three patterns. We illustrate their presentation by examples where terms are written in capital letters, functional words belong to the set {"a" "any" "for" "in" "is" "may" "of" "or" "the" "this" "to"}, there is one lexical scheme: "of", and six terms constitute the bootstrap: "WOOD" "COLOUR" "BEECH" "TIMBER", "DIESEL", "ENGINE".
1. Two terms (or candidate-terms) of the bootstrap appear in the same window
   A candidate-term is qualified when two existing terms (or candidate-terms) of the bootstrap appear frequently with almost the same arrangement. The most frequent arrangement becomes a candidate-term
   
   Example :
   
   ... *"the" "DIESEL", "ENGINE" "is"* ...
   ... *"this" "DIESEL", "ENGINE" "has"* ...
   ... *"a" "DIESEL", "ENGINE" "never"* ...
   
   The study of these three contexts of "DIESEL" and "ENGINE" will qualify "DIESEL ENGINE" as a candidate-term.
2. A term (or candidate-term) of the bootstrap and a lexical scheme word appear in the same window
   A candidate-term is qualified when a term (or candidate-term) of the bootstrap appears frequently with a lexical scheme word and with a word. This word then becomes a new candidate-term.
   
   Example :
   
   ... *"any" "shade" "of" "WOOD" "could"* ...
   ... *"this" "shade" "of" "WOOD" "is"* ...
   ... *"the" "shade" "of" "BEECH" "may"* ...
   ... *"new" "shade" "of" "TIMBER"*...
   ... *"same" "shade" "of" "WOOD" "in"* ...
   
   The study of these contexts will qualify "SHADE" as candidate-term.
3. A term (or candidate-terms) of the bootstrap without any lexical scheme nor second term
   A candidate-term is qualified when a term (or candidate-terms) of the bootstrap appears frequently with a word without any lexical scheme word nor second term.. This word then becomes a new candidate-term.
   
   Example :
   
   ... "use" "any" "soft" "WOODS" "to" "make" "this" ...
   ... "buy" "this" "soft" "WOODS" "or" "plastic" "for" ...
   ... "cheapest" "soft" "WOODS" "comes" "from" ...
   
   The study of these contexts will qualify "SOFT WOODS" as a candidate-term.

Thus, the ANA system is capable of extracting single-word terms and multi-word terms. It can also enrich an existing terminology.

The adaptiveness of the system to the corpus (and to the language of the corpus) is improved by the possible inference of the lexical scheme words as has been suggested in previous work (ref. 11).

## 2. TERM RECOGNITION

In a corpus, terms do not always appear with the reference form. For instance, we can meet some slight variations due to diacritic signs (accent, hyphenation) or in the plural, but also some deep variations are not so rare as the transformation of a compound word into a verbal form.

It is now admitted that variations represent 25% of the occurrences of terms. It is especially important to identify them for indexing, or for text mining.

Surprisingly, research especially dedicated to this area had been rare over the last 20 years. We can cite Jacquemin who worked specifically on the recognition of terms and their variations and built the FASTR system based on a fine linguistic study (ref. 12). In English, in addition to common linguistic variations (singular/plural for instance), four types of term variations can be recognized: coordinations (*recognition of simple and complex terms* is the coordination of *recognition of simple terms* and *recognition of complex terms*), modifications/substitutions (*book on heart malformations* and *book on malformations*), permutations (*blood flow* or *flow of blood*), and morphosyntactic variations (*colourless eyes*, *eyes without colour*). .

The FASTR system uses the formalism of unification grammar in which lexical rules are composed of lists of constraints on features and of a structure on categories which is out of context. The rules impose some constraints that must be respected, for example, genre and number. In addition to this basic formalism, meta-rules allow the manipulation of the lexical rules.

### Flexible-equality of terms

The flexible-equality of terms (ref. 13) is a mathematical operator which determines if two terms should be considered as equivalent. This operator uses a list of functional words (empty words) and depends on a parameter : k.

### 1. Flexible-equality of strings

The flexible-equality of two strings w and w' is based on the calculus of the minimum editing distance (ref. 14).

$$\text{dist}(w_{1,i}, w'_{1,i}) = \min (\ \text{dist}(w_{1,i-1}, w'_{1,i}) + q,$$
$$\text{dist}(w_{1,i}, w'_{1,j-1}) + q,$$
$$\text{dist}(w_{1,i-1}, w'_{1,j-1}) + p \cdot \text{dist}(w_{i,i}, w'_{j,j}) \ )$$

with • $w_{n,m}$ being the sub-string beginning at the nth letter and finishing after the mth letter of the word w.

• $\text{dist}(x, y) = 1 \quad$ if x = y
$\qquad\qquad\ \ = 0 \qquad$ otherwise

• q : cost of the insertion / deletion of one letter,

• p : cost of the substitution of one letter by another.

Generally, a substitution is considered as a deletion followed by an insertion, thus p=2q.

This algorithm has a high complexity: $\Theta(n^3)$. It can be easily transformed into another algorithm with a $\Theta(n^2)$ complexity by using a vector to memorize intermediary results (ref. 15).

We define the weighted distance (WD) of two words w and w' as the minimum editing distance weighted by the sum of the length of the two words.

$$\text{WD}(w, w') = \text{dist}(w, w') / (|w| + |w'|)$$

with • |w| the length of the word w (in number of letters)

- dist(w, w') = dist($w_{1,|w|}$, $w'_{1,|w'|}$)

This distance varies from 0, when words are strictly equal to, 1, when words are completely different.

We define the flexible-equality of strings (noted |~) as follows:

    w |~ w' <=> WD(w, w') ≤ 1 / k

        with  1/k being a threshold to be determined

1/k represents the proportion of variation acceptable in a word.

## 2. Flexible-equality of complex terms

The flexible-equality of strings cannot be used directly on compound words because of the high complexity of the algorithm. Thus we define an operator for this compound words.

This operator requires the succession of 3 steps :

### 1. Segmentation of the strings into words

The segmentation of a string into words is processed through the definition of a list of symbols which can be encountered in a string. This list includes all the letters of the considered alphabet (minuscules, majuscules and accentuated ones), the digits, and sometimes the minus charactere '-'.

A term X is defined as an ordered list of words $x_i$ : X = $x_{i,n}$.

    example:

    "*colour of a hammer*" is considered as the ordered list ("*colour*", "*of*", "*a*", "*hammer*")

### 2. Deletion of the functional words

Functional words (empty words) are defined in the list. In English we would find in this list articles ("a", "the", "this", etc.), prepositions ("of", "through", etc.), pronouns ("they", "who", etc.), etc.

We call restriction of a string X, and note R(X), the ordered list of words of X which are not functional words.

    example:

    X = "*colour of a hammer*"  R(X) = ("*colour*", "*hammer*")

                            "of" and "a" being functional words.

### 3. Respective comparisons of the words

Two strings are flexible-equal (noted '‖~') if the words of their restriction are flexible-equal.

    example: Here there is an m forgotten in hamer in the string Y

    X = "colour of a hammer"  R(X) = ("colour", "hammer")

    Y = "colour of any hamer"  R(Y) = ("colour", "hamer")

    ("colour" |~ "colour" and "hammer" |~ "hamer")     ⇒ (X ‖~ Y)

We define the distance (D) between two terms X and Y as the average of the distances between their restriction, words being compared respectively to their rank. This definition is useful when several candidates are flexible-equal. In such a case the one with the smallest distance will be chosen.

# 3. TERM ORGANISATION

Terminology structuring provides classes or links between terms. Classes are produced by clustering techniques based on similar word contexts (which describe what words are likely to be found in the immediate vicinity of a given word) (ref. 16) or similar distributional contexts (which show what words share the same syntactic environments) (ref. 17). Links result from automatic acquisition of relevant predicative or discursive patterns (ref. 10, 18, 19). Predicative patterns yield predicative relations such as <<cause>> or <<effect>> (ref. 20) whereas discursive patterns yield no-predicative relations such as hypernym (ref. 21) or synonymy links (ref. 22). Lastly, some tools exploit such automatic collected data for the purpose of automatic indexing (ref. 23) or information extraction (ref. 24).

In the field of terminology structuring, the main direction of research is the semantic classification from distributional analysis. Generally speaking, these methods are robust to extract classes between words, but have some disadvantages:

1. Clusters obtained with such techniques are not a priori significant (these clusters must be explained).
2. Clusters contain heterogeneous linguistic entities (for instance, a word and its antonym can be included in the same clusters).
3. Conceptual similarity, as well as semantic proximity, are "neutral" links which does not yield links labelled by semantic predicates.

Conversely, links between words that result from automatic acquisition of relevant predicative or discursive patterns are fine and accurate, but the acquisition of these patterns is a tedious task that requires substantial manual work. For our work, we use a low-cost system (called: Promethee) which extracts and uses lexico-syntactic patterns to acquire semantic relations between terms.

## The Promethee system

The Promethee system, for corpus-based information extraction, has two functionalities:
1. A corpus-based acquisition of lexico-syntactic patterns with respect to a specific conceptual relation.
2. The extraction of pairs of conceptually related terms through a database of lexico-syntactic patterns.

These functionalities are implemented in three main modules:

### Lexical Preprocessor

The lexical pre-processor receives raw texts as input. First, the text is tokenized (recognition of word and sentence boundaries), tagged and lemmatized. Noun phrases, acronyms and sequences of noun phrases are detected through regular expressions. The output of the lexical pre-processor is an enriched text with SGML tags.

### Shallow Parser and Classifier

This module extracts lexico-syntactic patterns relative to semantic relationships. This phase has been inspired by the works of P. Hearst (ref. 18) and implemented by a shallow parser associated with a classifier.

### Information Extractor

The information extractor acquires pairs of conceptually related terms by using a database of lexico-syntactic patterns. This database can be the output of the shallow parser and classifier, or may be manually specified.

The shallow parser is complemented with a classifier for the purpose of discovering new patterns through corpus exploration. This procedure is inspired by P. Hearst (ref. 18) and consists of the following seven steps:
1. Manually select a representative conceptual relation, e.g. the hypernym relation.
2. Collect a list of pairs of terms linked by the previous relation. This list of pairs of terms can be extracted from a thesaurus, a knowledge base or may be manually specified. For instance, the hypernym relation « *neocortex is-a-kind-of vulnerable area* » is used.
3. Find sentences in which conceptually related lemmatized terms occur. These sentences are lemmatized, and noun phrases are identified. They are represented as lexico-syntactic expressions. For instance, the previous relation *HYPERNYM* (*vulnerable area,neocortex* ) is used to extract the sentence: *Neuronal damage was found in the selectively <u>vulnerable areas</u> such as <u>neocortex,</u> striatum, hippocampus and thalamus* from a medical corpus.The sentence is then transformed into the following lexico-syntactic expression[1]: NP find in NP such as LIST
4. Find a common environment that generalizes the lexico-syntactic expressions extracted at the third step. This environment is calculated with the help of a function of similarity and a procedure of generalization that produce candidate lexico-syntactic patterns. For instance, from the previous

---

[1] NP stands for a noun phrase, and LIST for a succession of noun phrases

expression, and at least one other similar one, the following candidate lexico-syntactic pattern is deduced: NP such as LIST

5. Validate candidate lexico-syntactic patterns by an expert.
6. Use these validated patterns to extract additional candidate pairs of terms.
7. Validate candidate pairs of terms by an expert, and go to step 3.

At this level, two significant points make our technique different from Hearst's methodology:

1. A common environment relative to a set of sentences is extracted automatically by the former method and manually by Hearst.
2. The expert evaluation of candidate lexico-syntactic patterns or pairs of terms is absolutely necessary since all candidate patterns or pairs of terms do not denote the target relationship. This evaluation is not mentioned by Hearst.

The Promethee system has been used to extract discursive patterns relative to (1) a generic relation (hyponymy relation from a 1.3-million word French agricultural corpus and a 1.56-million word English medical corpus) and (2) two specific relations (merge and produce relations from a 0.9-million word English news stories corpus). Links extracted by the promethee system have a high precision, but cover only a part of targeted links occurring in the corpus.

# 4. FUTURE DEVELOPMENTS

## 4.1. Multilinguality

ACABIT, ANA and Promethee were designed for French, their adaptation to other languages mainly depends on the linguistic features of the targeted language. Previous experiments showed that this adaptation should be easy for non-agglutinative indo-european languages. For instance the adaptation to English language requires very slight adaptations, such as the new definition of base terms for ACABIT, new lists of functional words and lexical scheme words for ANA, Promethee would need more adaptations, for instance a tagger, a lemmatizer anda a noun phrase extractor.

The treatment of languages not belonging to Roman languages is also possible: we have customised ACABIT and ANA systems for Malagazy texts in previous work (ref. 25). The Malagasy language is very close to the Indonesian language. It has been written since the 19th century with the latin alphabet. Term identification from Malagazy texts presents some difficulties due to some of the linguistic characteristics of this language, for example the combinatorial variation, the agglutination of possessive and personal pronouns, the non-formal categories of the adverbs, and the homonymy between functional words and semantic words. However we succeeded in defining pre-treatments which allowed the utilization of both systems.

The adaptation of the flexible-equality operator will depend on the morphological feature of the considered language.

## 4.2. Acceptation of new formats

*XML format*
ACABIT accepts texts designed with an XML format and also produces terms designated by an XML format. This capability could be added to ANA and to the supple_equality operator.

*Unicode*
Unicode offers the possibility to code any symbol used by any language in the world. The adaptation of our systems to this standard is crucial for their multilingual future.

# Conclusion

Constituting terminologies of different technical or scientific domains and in different languages is crucial because it allows people to conceptualize and understand techniques and sciences in their own languages. These linguistic resources represent a major contribution to the increase of the education level of a population.

Nevertheless Terminology processing is gaining a real place in Natural Language Processing technologies.

Nervertheless, this linguistic effort is too often not processed, essentially because of its costs. The last decade saw the appearance of new tools, destined to help in the collecting, retrieving and structuring terms. Recent advances in the normalisation of linguistic symbols (with Unicode), and linguistic study of the targeted languages features should allow the adaptation of these existing tools to different languages.